\documentclass[copyright,creativecommons]{eptcs}
\providecommand{\event}{ACL2 Workshop 2022} 
\usepackage{underscore}           

\usepackage{listings}

\usepackage{float}

\newcommand{\TinySkip}{\vspace{1ex}\noindent}

\newcommand{\rbx}[1]{\mbox{#1}}

\title{Modeling Asymptotic Complexity Using ACL2}
\author{\large William D. Young, University of Texas at Austin}

\author{William D. Young
\institute{Department of Computer Science\\
University of Texas at Austin\\
Austin, Texas}
\email{byoung@cs.utexas.edu}
}
\def\titlerunning{Modeling Asymptotic Complexity}
\def\authorrunning{William D. Young}
\begin{document}
\maketitle

\begin{abstract}
  The theory of asymptotic complexity provides an approach to
  characterizing the behavior of programs in terms of bounds on the
  number of computational steps executed or use of computational
  resources.  We describe work using ACL2 to prove complexity
  properties of programs implemented in a simple imperative
  programming language embedding via an operational semantics in ACL2.
  We simultaneously prove functional properties of a program and its
  complexity.  We illustrate our approach by describing proofs about a
  binary search algorithm, proving both that it implements binary
  search on a sorted list and that it is $O(\log_2(n))$, where $n$ is
  the length of the list.
\end{abstract}

\section{Introduction}
\label{Introduction}

The theory of asymptotic complexity provides a systematic approach to
characterizing the limiting behavior of a function as its argument
tends toward infinity. A family of notations, collectively called {\it
  Bachmann-Landau} or asymptotic notations, allow characterizing upper
bounds, lower bounds, and tight bounds of one function in terms of
another function.  These have been used in analytic number theory for
more than a century, but became ubiquitous in computing following the
groundbreaking work of Hartmanis and
Stearns\cite{HartmanisStearns}. These notations provide convenient
abstract characterizations of the difficulty of computing specific
algorithms, independent of any particular hardware platform.

The most common asymptotic notation is the big-O notation for
estimating an upper bound on the time or space complexity of an
algorithm.  It can be defined as follows\cite{Sipser}:

\TinySkip
\begin{changemargin}{20pt}{20pt}
\noindent{\bf Definition:} Let $f$ and $g$ be functions $f, g: N \rightarrow
R^+$.  We say that $f(n) = O(g(n))$ if positive integers $c$ and $n_0$
exist such that for every integer $n \ge n_0$,
$$f(n) \le c\cdot g(n).$$
\end{changemargin}

\TinySkip
\noindent When $f(n) = O(g(n))$ we say that $g(n)$ is an {\it upper
  bound} for $f(n))$, or more precisely, that $g(n)$ is an {\it
  asymptotic upper bound} for $f(n)$, to emphasize that constant
factors are suppressed.  

As an example, if $f(n) = 3n^3 + 7n^2 - 5$, then we can easily prove
that $f(n) = O(n^3)$, choosing $c = 4$ and $n_0 = 7$ (though
infinitely many other choices of $c$ and $n_0$ would work as well). In
the context of computational complexity, the function being bounded
typically describes the number of computational steps or the number of
units of storage required in execution of an algorithm.  Thus, for
example, we might say that ``binary search is $O(\log_2(n))$,'' where
$n$ is the length of the sorted list being searched, for the following
reason: Executing the algorithm takes no more than $c\cdot \log_2(n)$
steps whenever $n \ge n_0$, for some positive integers $c,\, n_0$.
The proof entails exhibiting specific witnesses for $c$ and $n_0$.

The goal of the research described in this paper is to formalize and
prove big-O properties of algorithms using ACL2.  Since the definition
of big-O is logically higher-order, we make certain concessions to the
first-order nature of the ACL2 logic.  Specifically, we proved big-O
properties of specific algorithms, without defining big-O abstractly.
For example, we proved that an implementation of binary sort has
complexity bounded by $\log_2(n)$, where $n$ is the length of the list
being searched.\footnote{We also carried out a proof that linear
  search has linear complexity and other proofs of the correctness of
  some specific algorithms coded in our iterative language.  These
  proofs are not described here.}  To do this we defined {\it
  logarithmic complexity} as a first-order predicate rather than as an
instance of the higher-order function $O(-)$.

We approached this by defining a simple imperative language and an
interpreter for that language within ACL2.  The interpreter defines a
traditional operational semantics for programs within the language,
using a clock argument to guarantee termination.  We prove functional
properties of our programs using the interpreter semantics in the
standard way.  The interpreter also keeps a count of computational
steps as execution proceeds.  Then we prove for specific programs in
our language that the count of steps taken during execution of the
program is bounded (in the big-O sense) by another specific function.
As an example, we prove that a program in our language that implements
binary search on a sorted list of length $n$, is
$O(\log_2(n))$.\footnote{We used a function that approximates
  $\log_2(n)$ by counting the number of times $n$ can be divided in
  half, using integer division.  One reviewer pointed out that there
  are at least two definitions of {\tt ilog} in community books.  Our
  definition is off by 1 from these, but otherwise provably
  equivalent.}

The paper is structured as follows.  The following section cites some
related work.  Section \ref{The Language} describes our simple
imperative language and its interpreter semantics.  In Section
\ref{Proving Correctness and big-O}, we outline our approach to
proving simultaneously the functional correctness of a program and its
computational complexity.  This is illustrated by the proof of a
binary search algorithm on a sorted list.  We prove that the algorithm
has complexity of $O(\log_2(n))$, where $n$ is the length of the list.
Finally, in Section \ref{Conclusions} we draw some conclusions and
point to potential extensions of this work.

\section{Related Work}
\label{Related Work}

The notion of embedding a language within the Boyer-Moore family of
logics, of which ACL2 is the latest instantiation, has a very long
history\cite{BoyerMoore96} that includes modeling
software\cite{Moore99}, hardware\cite{BrockKaufmannMoore96}, and
systems\cite{BevierHuntMooreYoung89}.  See the ACL2 homepage
publications list for a
selection.\footnote{\url{https://www.cs.utexas.edu/users/moore/publications/acl2-papers.html.}}
We are unaware of any work with ACL2 specifically aimed at modeling
the asymptotic complexity of algorithms, though Boyer and Moore
certainly did careful analysis of the complexity of algorithms they
developed such as their fast string search
algorithm\cite{BoyerMoore77} and fast majority vote
algorithm\cite{BoyerMoore91}.

However, research on modeling and proving asymptotic complexity
results has been done using other automated formal reasoning systems.
That work encompasses three distinct area of research: expressing
within a mechanized formal logic big-O and related
notions,\footnote{Others notations in this family include Omega (lower bound), Theta (exact
  bound), and little-O (upper bound excluding exact bound).}
extracting complexity results automatically from programs, and proving
complexity results about specific algorithms/programs.


Exemplifying the first approarch is the work of Iqbal et
al.\cite{Iqbal19} who formalized big-O and related notions in the
HOL-4 logic and checked proofs of classical properties such as
transitivity, symmetry and reflexivity using the HOL-4 prover.  They
do not appear to have used the definitions in the analysis of any
specific programs.  Most other efforts in this space have gone beyond
merely modeling and manipulating notations to analyzing algorithms or
programs.  Modeling and verification tools like HOL and Coq that
implement reasoning for higher-order logic allow modeling big-O and
related notions, which are inherently higher order, more directly than
does ACL2.

Inferring the complexity of algorithms is theoretically undecidable,
but possible for many programs.  Steps in that direction (for
worst-case analysis) were described by Wegbreit\cite{Wegbreit}, Le
M\'etayer\cite{LeMetayer}, and Benzinger\cite{Benzinger}.
Average-case analysis is somewhat harder since it requires a
distribution function for the data and probabilistic reasoning.
Hickey and Cohen\cite{HickeyCohen} describe a performance compiler
that generates recurrence relations characterizing the average-case
complexity of programs.  Similar work is described by Flajolet et
al.\cite{Flajolet}

Amortized analysis aims to model average case behavior over a suite of
operations, where some may be considerably more expensive than the
average.  Hofmann and Jost\cite{HofmannJost} applied automatic
type-based amortized analysis to heap usage of functional programs.
Carbonneaux et al.\cite{Carbonneaux} generate Coq proof objects to
certify inferred resource bounds.  Nipkow and
Brinkop\cite{NipkowBrinkop} have formalized in Isabelle/Hol a
framework for the analysis of amortized complexity of functional data
structures.

Perhaps closest to our work is that of
Gu\'eneau\cite{Gueneau18,Gueneau20}. Using and extending an existing
program verification framework, CFML\cite{Chargueraud}, Gu\'enaeu uses
Separation Logic with Time Credits to model big-O and its related
notions and prove the asymptotic complexity of OCaml programs using
the Coq automated reasoning system.  A very extensive bibliography on
research in modeling and proving asymptotic complexity results with
automated reasoning systems is available in Gu\'enau's
dissertation\cite{Gueneau20}.





\section{A Simple Imperative Programming Language}
\label{The Language}

Our primary goal is to prove big-O complexity results for specific
algorithms using ACL2.  To do so, we define a simple imperative
programming language embedded in the ACL2 logic.

\subsection{Expression Sublanguage}
\label{Expression Sublanguage}

\begin{figure}[hbtp]
\begin{tabbing}
\hspace*{1.25in}\=\hspace*{2.75in}\=\kill
\>$(\rbx{var}\; s)$ \>variable\\
\>$(\rbx{lit}\; .\; v)$ \>literal\\
\>$(\rbx{lit}\; .\; (v_0,\, \ldots\, v_k))$ \>list literal\\
\>$(==\; e_1\; e_2)$  \>equals\\
\>$(+\; e_1\; e_2)$ \>addition\\
\>$(-\; e_1\; e_2)$ \>subtraction\\
\>$(*\; e_1\; e_2)$ \>multiplication\\
\>$(//\; e_1\; e_2)$ \>integer division\\
\>$(<\; e_1\; e_2)$ \>less than \\
\>$(<=\; e_1\; e_2)$ \>less than or equal\\
\>$(>\; e_1\; e_2)$ \>greater than\\
\>$(>=\; e_1\; e_2)$ \>greater than or equal\\
\>$(\rbx{len}\; e)$\>len of a list\\
\>$(\rbx{ind}\; e_1\; e_2)$ \>index into a list.
\end{tabbing}
    \caption{Expression Language}
    \label{expression-language}
\end{figure}

Our language includes an expression sublanguage as shown in Figure
\ref{expression-language}.  Here $s$ denotes a symbol, $v$ either a
number or symbol, and $e$ an expression within this sublanguage.  The
semantics of our expression sublanguage is defined by an evaluator
function {\tt (exeval x status vars)}.  Here {\tt x} is the expression
to be evaluated, {\tt status} is a boolean indicating whether
evaluation should proceed and {\tt vars} is a variable alist
associating variable names to values (symbols, numbers, or lists of
values).  Expression evaluation returns a triple {\tt (status, value,
  steps)}, where {\tt status} is a boolean indicating whether or not
the evaluation raised an error, {\tt value} is the result of the
evaluation (assuming no error occurred), and {\tt steps} records the
number of steps taken by the evaluation.  In Figure \ref{exeval} we
exhibit part of the definition of {\tt exeval};  clauses omitted are
similar to those shown.

\begin{figure}[hbtp]
\lstset{basicstyle=\ttfamily\small}
\lstset{commentstyle=\small\texttt}
\lstset{aboveskip=\baselineskip,
        belowskip=1.2\baselineskip}
\begin{lstlisting}[language=Lisp,frame=single]
(defun exeval (x status vars)
  (if (not status)
      ; Execution proceeds only if status is T
      (exeval-error)
    ; No legal expression is an atom
    (if (atom x)
        (exeval-error)
      (case (operator x)
            ; Variables
            (var (if (definedp (param1 x) vars)
                     (mv t (lookup (param1 x) vars) 1)
                   (exeval-error)))
            ; Literals
            (lit (if (valuep (cdr x))
                     (mv t (cdr x) 1)
                   (exeval-error)))
            ; Addition
            (+ (mv-let (stat1 val1 steps1)
                       (exeval (param1 x) status vars)
                       (mv-let (stat2 val2 steps2)
                               (exeval (param2 x) status vars)
                               (if (and stat1 stat2
                                        (acl2-numberp val1)
                                        (acl2-numberp val2))
                                   (mv t (+ val1 val2) 
                                       (+ 1 steps1 steps2))
                                 (exeval-error)))))
               ...

            ; Index into a list (ind i lst)
            (ind (mv-let (stat1 val1 steps1)
                         (exeval (param1 x) status vars)
                         (mv-let (stat2 val2 steps2)
                                 (exeval (param2 x) status vars)
                                 (if (and stat1 stat2
                                          (natp val1)
                                          (listp val2)
                                          (< val1 (len val2)))
                                     (mv t (nth val1 val2) 
                                         (+ 1 steps1 steps2))
                                   (exeval-error)))))
            (otherwise (exeval-error))))))


    \end{lstlisting}
    \caption{Part of the Definition of {\tt exeval}}
    \label{exeval}
\end{figure}

Note that arguments within any subexpression are ``type checked'' to
ensure that evaluation will succeed.  Thus, arithmetic and relational
functions are applied only to numbers; {\tt len} is applied only to
lists; etc.  It would be straightforward to relax some of the
restrictions, e.g., comparing symbols alphabetically.  
Also, evaluation is strict in error detection; an error at any level
propogates to the top.  The presumption is that we only care about the
result and the complexity of the computation if it terminates
normally, without error.  

The number of steps assessed for any evaluation is somewhat
arbitrary.  For example, we can perform the following computation:
\lstset{commentstyle=\small\texttt}
\lstset{aboveskip=\baselineskip,
        belowskip=1.2\baselineskip}
\begin{lstlisting}[language=Lisp,frame=single] 
ACL2 !>(exeval '(+ (var  x) (* (var  y) (lit . 10)))    ; expression
                t                                       ; status
                '((x . 7) (y . 5)))                     ; alist
(T 57 5)
\end{lstlisting}
The resulting triple indicates that the evaluation proceeded without
error and returned a value of 57 in 5 steps (1 each to evaluate two
variables and one literal, and 1 each for the multiplication and
addition).  We could easily adjust the numbers of steps assessed for
each operation, e.g., make multiplication more expensive than
addition.  Such changes would not affect the complexity of any
programs, as long as these operations remained $O(1)$.  We also assume
that both indexing into a list ({\tt ind}) and computing the length of
a list ({\tt len}) are $O(1)$ operations.  Again, that could easily be
adjusted, though treating {\tt len} as an $O(n)$ operation, as it
might be in Lisp, could change the big-O complexity of programs using
it.

\subsection{Statement Sublanguage}
\label{Statement Sublanguage}

\begin{tabbing}
\hspace*{0.5in}\=\hspace*{2.0in}\=\kill
\>$(\rbx{skip})$ \>no-op  \\
\>$(\rbx{assign}\; v\; e)$    \>assign expression e to variable v  \\
\>$(\rbx{return}\; e)$    \>assign expression e to variable {\tt 'return}   \\
\>$(\rbx{if-else}\; e\; s_1\; s_2)$    \>if expression e, do statement
$s_1$, else do $s_2$\\
\>$(\rbx{while}\; e\; s)$    \>while expression e is true, do statement $s$  \\
\>$(\rbx{seq}\; s_1\; s_2)$    \>  do statement $s_1$, then $s_2$ \\
\end{tabbing}
\vspace*{-1.0ex}

Our simple imperative language contains the six types of statements
shown above. 
We provide an operational semantics for this language in the form of
an interpreter: {\tt (run stmt status vars steps count)} Here {\tt
  stmt} is a statement to execute, {\tt status} a symbol indicating
whether the computation should continue, {\tt vars} a variable alist,
{\tt steps} a tally of the steps so far, and {\tt count} our ``clock''
argument to guarantee that execution terminates.  Function {\tt run}
returns a triple {\tt (status, vars, steps)}.  Unlike the expression
sublanguage, here {\tt status} is not a boolean, but a symbol.  A
value of {\tt 'OK} indicates that execution should continue; {\tt
  'RETURNED} indicates that a result was assigned to variable {\tt
  result}; any other value signals an error, typically {\tt 'ERROR} or
{\tt 'TIMED-OUT} (indicated that {\tt count} was insufficient for the
recursive depth of the call tree).  Note that it is possible in ACL2
to define an interpreter for this language without the clock argument.
Several approaches are described in \cite{CowlesGreveYoung}.

\begin{figure}[hbtp]
\lstset{basicstyle=\ttfamily\small}
\lstset{commentstyle=\small\texttt}
\lstset{aboveskip=\baselineskip,
        belowskip=1.2\baselineskip}
\begin{lstlisting}[language=Lisp,frame=single]
(defun run (stmt status vars steps count)
  (declare (xargs :measure (two-nats-measure count (acl2-count stmt))))
  (if (not (okp status))
      (mv status vars steps)
    (if (zp count)
        (mv 'timed-out vars steps)
      (case (operator stmt)
            (skip   (run-skip stmt vars steps))
            (assign (run-assignment stmt vars steps))
            (return (run-return stmt vars steps))
            (seq    (mv-let (new-stat new-vars steps1)
                            (run (param1 stmt) status vars steps count)
                            (run (param2 stmt) new-stat 
                                 new-vars steps1 count)))
            (if-else (mv-let (eval-stat eval-val eval-steps)
                             (exeval (param1 stmt) t vars)
                             (if (not eval-stat)
                                 (run-error vars)
                               (if eval-val
                                   (run (param2 stmt) status vars 
                                        (+ 1 steps eval-steps) count)
                                 (run (param3 stmt) status vars 
                                      (+ 1 steps eval-steps) count)))))
            (while (mv-let (test-stat test-val test-steps)
                           (exeval (param1 stmt) t vars)
                           (if (not test-stat)
                               (run-error vars)
                             (if test-val
                                 (mv-let (body-stat body-vars body-steps)
                                         (run (param2 stmt) status vars
                                              (+ 1 steps test-steps) 
                                              count)
                                         (run stmt body-stat body-vars 
                                                   body-steps 
                                                   (1- count)))
                               (mv 'ok vars (+ 1 test-steps steps))))))
            (otherwise (run-error vars))))))
    \end{lstlisting}
    \caption{The Interpreter for our Imperative Language}
    \label{interpreter}
\end{figure}
The definition of the interpreter is in Figure \ref{interpreter}.
Here, {\tt run-skip}, {\tt run-return} and {\tt run-assignment} are
non-recursive functions which evaluate {\tt skip}, {\tt return} and
{\tt assign} statements, respectively.  These are shown in Figure
\ref{run-assign-return}.  These don't require the {\tt status}
argument since they're never called unless {\tt status} is {\tt 'OK};
they don't require {\tt count} since they are non-recursive.
\begin{figure}[hbtp]
\lstset{basicstyle=\ttfamily\small}
\lstset{commentstyle=\small\texttt}
\lstset{aboveskip=\baselineskip,
        belowskip=1.2\baselineskip}
\begin{lstlisting}[language=Lisp,frame=single]
(defun run-skip (stmt vars steps)
  ; Statement is (skip)
  (declare (ignore stmt))
  (mv 'ok vars steps))

(defun run-assignment (stmt vars steps)
  ; Statement has form (assign var expression)
  (mv-let (eval-stat val eval-steps)
          (exeval (param2 stmt) t vars)
          ; lhs must be a variable
          (if (and (varp (param1 stmt))
                   eval-stat)
              (mv 'ok 
                  (store (cadr (cadr stmt)) val vars) 
                  (+ 1 eval-steps steps))
            (run-error vars))))

(defun run-return (stmt vars steps)
  ; Statement has form (return expression)
  (mv-let (eval-stat val eval-steps)
          (exeval (param1 stmt) t vars)
          (if eval-stat
              (mv 'returned 
                  (store 'result val vars) 
                  (+ 1 eval-steps steps))
            (run-error vars))))
    \end{lstlisting}
    \caption{Run, Assign, and Return}
    \label{run-assign-return}
\end{figure}

As with expression evaluation, all arguments are ``type checked'' and
error checking is strict.  All of our big-O results assume that
execution of the program terminates without error.  Again, the
counting of steps is somewhat arbitrary, but the choices seem to us to
be defensible.\footnote{We note, for example, that similar analysis by
  Gu\'enau, et al.\cite{Gueneau20} assesses one time unit for calling a
  function or entering a loop body, and zero units for other
  operations.}  For example, running the statement:
\begin{center}
{\tt (if-else test true-branch false-branch)} 
\end{center}
requires one plus the number of steps to evaluate the test plus the
number of steps to execute {\tt true-branch} or {\tt false-branch}, as
appropriate.  These easily could be adjusted.

\subsection{Programs in the Language}
\label{Programs in the Language}

Our simple language is almost certainly Turing complete, though we
haven't proved that.  We define programs in the language as ACL2
literals.  To see how it works, consider the following Python program
that computes binary search.
\lstset{basicstyle=\ttfamily\small}
\lstset{commentstyle=\small\texttt}
\lstset{aboveskip=\baselineskip,
        belowskip=1.2\baselineskip}
\begin{lstlisting}[language=Lisp,frame=single]
def BinarySearch( key, lst ):
    low = 0
    high = len(lst) - 1
    while (high >= low):
        mid = (low + high) // 2
        if key == lst[mid]:
            return mid
        elif key < lst[mid]:
            high = mid - 1
        else:
            low = mid + 1
    return -1
\end{lstlisting}

In Figure \ref{binarysearch} is the version of this program in our simple imperative
language.\footnote{Here {\tt (seqn x y ... w z)} is an abbreviation for
  {\tt (seq x (seq y ... (seq w z))...)}.}
It's pretty easy to see that this is a straightforward translation of
the Python program. This translation is likely not difficult to
automate, though we haven't tried to do so.  Performing any sort of
optimizations automatically would be more challenging.

\begin{figure}[hbtp]
\lstset{basicstyle=\ttfamily\small}
\lstset{commentstyle=\small\texttt}
\lstset{aboveskip=\baselineskip,
        belowskip=1.2\baselineskip}
\begin{lstlisting}[language=Lisp,frame=single]
(defun binarysearch (key lst)
  `(seqn (assign (var low) (lit . 0))
         (assign (var high) (- (len ,lst) (lit . 1)))
         (while (<= (var low) (var high))
           (seq (assign (var mid) 
                        (// (+ (var low) (var high)) (lit . 2)))
                (if-else (== ,key (ind (var mid) ,lst))
                         (return (var mid))
                         (if-else (< ,key (ind (var mid) ,lst))
                                  (assign (var high) 
                                          (- (var mid) (lit . 1)))
                                  (assign (var low) 
                                          (+ (var mid) (lit . 1)))))))
         (return (lit . -1)))))
\end{lstlisting}
    \caption{Binary Search in Our Imperative Language}
    \label{binarysearch}
\end{figure}


We can run our program with concrete data by invoking the interpreter
on this constant with appropriate arguments. For example, 
\begin{center}
\begin{minipage}[t]{\linewidth}
\lstset{basicstyle=\ttfamily\small}
\lstset{commentstyle=\small\texttt}
\lstset{aboveskip=\baselineskip,
        belowskip=1.2\baselineskip}
\begin{lstlisting}[language=Lisp,frame=single]
ACL2 !>(run (binarysearch '(lit . 4) 
                          '(lit . (0 1 2 3 4 5 6 7)))
            'OK nil 0 10)
(RETURNED ((LOW . 4)
           (HIGH . 4)
           (MID . 4)
           (RESULT . 4))
          77)
\end{lstlisting}
\end{minipage}
\end{center}
This shows that the computation returned after 77 steps with an
updated variable alist, including the index/answer in variable {\tt
  RESULT}.  Note that several var/value pairs were added to the alist
during execution, corresponding to the procedure's locals and the
result value.  Below is an example using variables in the alist:
\lstset{basicstyle=\ttfamily\small}
\lstset{commentstyle=\small\texttt} \lstset{aboveskip=\baselineskip,
  belowskip=1.2\baselineskip}
\begin{lstlisting}[language=Lisp,frame=single]
ACL2 !>(run (binarysearch '(var key) '(var lst))
            'OK '((key . 4) (lst . (0 1 3 5 7 9 10))) 0 10)
(RETURNED ((KEY . 4)
           (LST 0 1 3 5 7 9 10)
           (LOW . 3)
           (HIGH . 2)
           (MID . 2)
           (RESULT . -1))
          91)
\end{lstlisting}
In this case, {\tt RESULT} contains -1, indicating that the search
failed after 91 steps.  Finally, here's the same computation, but with
an insufficient clock value: \lstset{basicstyle=\ttfamily\small}
\lstset{commentstyle=\small\texttt} \lstset{aboveskip=\baselineskip,
  belowskip=1.2\baselineskip}
\begin{lstlisting}[language=Lisp,frame=single]
ACL2 !>(run (binarysearch '(var key) '(var lst))
            'OK '((key . 4) (lst . (0 1 3 5 7 9 10))) 0 1)
(TIMED-OUT ((KEY . 4)
            (LST 0 1 3 5 7 9 10)
            (LOW . 0)
            (HIGH . 2)
            (MID . 3))
           33)
\end{lstlisting}
Our proofs about big-O properties of functions assume that the
computation does not raise an error or time out. 

\section{Proving Correctness and big-O}
\label{Proving Correctness and big-O}

Our proofs do more than verifying the big-O behavior of the program.
We simultaneously prove the functional correctness of the program.
This is not particularly novel, but assures that the program for which
we are establishing complexity results is also functionally correct.
We describe this aspect of our proof effort below in Section
\ref{Functional Correctness}.  The more novel aspect of our work is
verifying the big-O behavior of programs.  This is described in
Section \ref{Computational Complexity}.  


\subsection{Functional Correctness}
\label{Functional Correctness}

We characterize the functional behavior of a program by defining a
recursive functions that closely mimics the execution of our
imperative program.  For example, for the proof of our binary search
program, we defined a function {\tt (recursiveBS key lst)} that
behaves ``in the same way'' as our iterative version.  This is
illustrated in Figure \ref{recursiveBS}. 
Notice that the computation carries along the local parameters of the
Python computation as well as a count of the number of recursive
calls.  The locals are needed to show in the inductive proof that the
recursive and non-recursive versions coincide, since these variables
are all stored in the iterative state.  Our proof shows that, in each
step of the execution, the values of variables stored in the state are
exactly those computed by our recursive version.  The count of
recursive calls is needed because the time complexity of the
computation is ultimately a function of the number of recursive calls
made.

\begin{figure}[hbtp]
\lstset{basicstyle=\ttfamily\small}
\lstset{commentstyle=\small\texttt}
\lstset{aboveskip=\baselineskip,
        belowskip=1.2\baselineskip}
\begin{lstlisting}[language=Lisp,frame=single]
(defun recursiveBS-helper (key lst low mid high calls)
  ;; This performs a recursive binary search for key in 
  ;; lst[low..high].  It returns a 5-tuple (success low mid high calls).
  ;; We need all of those values to do the recursive proof.
  (declare (xargs :measure (nfix (1+ (- high low)))))
  (if (or (< high low)
	  (not (natp low))
	  (not (integerp high))
	  )
      (mv nil low mid high calls)
    (let ((newmid (floor (+ low high) 2)))
      (if (equal key (nth newmid lst))
	  (mv t low newmid high calls)
	(if (< key (nth newmid lst))
	    (recursiveBS-helper key lst low 
                                newmid (1- newmid) (1+ calls))
	  (recursiveBS-helper key lst (1+ newmid) 
                              newmid high (1+ calls)))))))

(defun recursiveBS (key lst)
  ;; This is the recursive version of binary search
  (mv-let (success low mid high calls)
	  (recursiveBS-helper key lst 0 nil (1- (len lst)) 0)
	  (declare (ignore low high calls))
	  (if success mid -1)))
\end{lstlisting}
\caption{Recursive Equivalent of our Binary Search Program}
\label{recursiveBS}
\end{figure}

As an example, we prove that if {\tt (member-equal keyval lstval)},
where {\tt keyval} and {\tt lstval} are values stored in the alist in
appropriate variables, then the following is true:
\lstset{basicstyle=\ttfamily\small}
\lstset{commentstyle=\small\texttt}
\lstset{aboveskip=\baselineskip,
        belowskip=1.2\baselineskip}
\begin{lstlisting}[language=Lisp,frame=single]
   (equal (run (binarysearch key lst) 'ok vars 0 count)
          (mv-let (success endlow endmid endhigh endcalls)
                  (recursivebs-helper keyval lstval 
                                      0 nil (1- (len lstval)) 0)
                  (mv 'returned
                      (store 'result endmid
                             (store 'mid endmid
                                    (store 'high endhigh
                                           (store 'low endlow vars))))
                      (+ 25 (* 26 endcalls)))))
\end{lstlisting}
This characterizes explicitly what state is computed by execution of
our iterative program; the values in the resulting variable alist are
precisely those computed by its recursive counterpart.  Notice that
the number of steps taken is a function of the computed number of
recursive calls.

Finally, we define a second, simpler recursive version {\tt
  recursiveBS2}, which omits the local variables, and prove that the two
recursive versions return the same ``result.''  By transitivity, the
iterative version and the simpler recursive version return the same
result.  Finally, we can see that our iterative program actually
computes binary search by examining our ``simpler'' recursive
function, shown here: 
\lstset{basicstyle=\ttfamily\small}
\lstset{commentstyle=\small\texttt} \lstset{aboveskip=\baselineskip,
  belowskip=1.2\baselineskip}
\begin{lstlisting}[language=Lisp,frame=single]
(defun recursiveBS2-helper (key lst low high)
  (if (or (< high low)
          (not (natp low))
          (not (integerp high))
          )
      -1
    (let ((newmid (floor (+ low high) 2)))
      (if (equal key (nth newmid lst))
          newmid
        (if (< key (nth newmid lst))
            (recursiveBS2-helper key lst low (1- newmid))
          (recursiveBS2-helper key lst (1+ newmid) high))))))

(defun recursiveBS2 (key lst)
  (recursiveBS2-helper key lst 0 (1- (len lst))))
\end{lstlisting}
Our iterative function computes binary search if this recursive
version does. Hence, it remains to show that this recursive function
actually searches a list.  We proved the following theorem:
\lstset{basicstyle=\ttfamily\small}
\lstset{commentstyle=\small\texttt} \lstset{aboveskip=\baselineskip,
  belowskip=1.2\baselineskip}
\begin{lstlisting}[language=Lisp,frame=single]
(defthm recursiveBS2-searches
  (implies (and (acl2-numberp key)
		(number-listp lst)
		(sorted lst))
	   (let ((index (recursiveBS2 key lst)))
	     (and (implies (member-equal key lst)
			   (equal (nth index lst) key))
		  (implies (not (member-equal key lst))
			   (equal index -1))))))
\end{lstlisting}

\subsection{Computational Complexity}
\label{Computational Complexity}

Since the interpreter for our simple imperative language counts
execution steps, it should be straightforward to characterize the
big-O behavior of the program in terms of the step count.  While
that's true in theory, it proved to be rather difficult in practice.
In this section, we illustrate the process of illustrating how we
prove that the count of the number of steps is bounded by our big-O
bounding function---in the case of binary search, $\log_2$.  

We defined an integer approximation of $\log_2(n)$; Assuming $n$ is a
positive integer, $\log_2(n)$ is approximately how often you can halve
$n$ before you reach $0$.  This suggests the following function:
\lstset{basicstyle=\ttfamily\small}
\lstset{commentstyle=\small\texttt} \lstset{aboveskip=\baselineskip,
  belowskip=1.2\baselineskip}
\begin{lstlisting}[language=Lisp,frame=single]
(defun log2 (n)
  (if (zp n)
      0
    (1+ (log2 (floor n 2)))))
\end{lstlisting}
This was the definition we used in proofs; notice it is well-suited for
reasoning about binary search.\footnote{Note that the base case
  probably should have been {\tt (or (zp n) (equal n 1))}. Without
  that the value is off by one for $n > 1$, but doesn't affect any of
  the complexity results. With that change, our version is equivalent
  to the definition of {\tt ilog} in at least two of the community
  books.}

Recall our definition of big-O from Section \ref{Introduction}.
Rather than try to define the intrinsically higher-order function
$O(-)$, we define an instance of it, {\it logarithmic complexity},
appropriate for complexity proofs about our binary search function.
In Figure \ref{code.2} is our rendering of this in ACL2. {\tt
  function-logarithmic1} defines the following predicate.  Suppose our
program (which is a quoted constant) is run on the interpreter with an
{\tt 'OK} initial status, variable alist, zero previous steps, and an
adequate clock.  Then the number of steps taken has the appropriate
relation to the two positive integer constants $c$ and $n_0$.  Here,
parameter {\tt log-of} specifies of which structure's size we're
taking the log.  {\tt function-logarithmic2} merely adds the
existential quantification with respect to variables $c$ and $n_0$.

\begin{figure}[hbtp]
\lstset{basicstyle=\ttfamily\small}
\lstset{commentstyle=\small\texttt}
\lstset{aboveskip=\baselineskip,
        belowskip=1.2\baselineskip}
\begin{lstlisting}[language=Lisp,frame=single]
(defun-sk function-logarithmic1 (program log-of c n0 vars count)
  ;; This says that program (which is just a literal) can be run
  ;; against the variable-alist vars with count.  It says that 
  ;; the number of steps taken to run the program are logarithmic
  ;; in the size of parameter log-of.  Params c and n0 are the two
  ;; variables of the definition of asymptotic complexity.
  (forall (n)
          (implies (and (equal n (len log-of))
                        (<= n0 n))
                   (mv-let (run-stat run-vars run-steps)
                           (run program 'ok vars 0 count)
                           (declare (ignore run-stat run-vars))
                           (and (<= 0 run-steps)
                                (<= run-steps (* c (log2 n))))))))

(defun-sk function-logarithmic2 (program log-of vars count)
  (exists (c n0)
          (and (posp c)
               (posp n0)
               (function-logarithmic1 program log-of c n0 vars count))))

    \end{lstlisting}
    \caption{Definition of Function Logarithmic}
    \label{code.2}
\end{figure}

Finally, we need to prove that our specific program satisfies this
predicate. This is the content of lemma {\tt
  binarysearch-logarithmic-lemma} shown in Figure \ref{code.1}.
This lemma asserts the following:  if we assume that
\begin{enumerate}
\item variable {\tt 'key} is associated with a number {\tt keyval} in our variable alist;
\item variable {\tt 'lst} is associated with a sorted list {\tt lstval} of numbers in our
  variable alist; and
\item the computation does not time-out,
\end{enumerate}
then our binary search function is bounded above, in the appropriate
big-O sense, by {\tt (log2 (len lstval))}.  The proof of this requires
supplying specific values of $c$ and $n_0$; we supplied
$c = 51; n_0 = 26$, though many other values would have worked as
well.  We have performed proofs showing that linear search is
linear in the length of the input list and similar results for several
other programs not described here.

\begin{figure}[hbtp]
\lstset{basicstyle=\ttfamily\small}
\lstset{commentstyle=\small\texttt}
\lstset{aboveskip=\baselineskip,
        belowskip=1.2\baselineskip}
\begin{lstlisting}[language=Lisp,frame=single]
(defthm binarysearch-logarithmic-lemma
  (let ((keyval (lookup 'key vars))
        (lstval (lookup 'lst vars)))
    (implies 
       (and (acl2-numberp keyval)
            (number-listp lstval)
            (sorted lstval)
            (integerp count)
            (not (timed-outp 
                    run-status (run (binarysearch '(var key) '(var lst)) 
                                                  'ok vars 0 count)))))
       (function-logarithmic2 (binarysearch '(var key) '(var lst))
                              (lookup 'lst vars) vars count))))
    \end{lstlisting}
    \caption{Binary Search Logarithmic}
    \label{code.1}
\end{figure}

\subsection{Subtleties of this Approach}
\label{Subtleties of this Approach}




Because we carefully count every step in the execution of our
program, we obtain a very fine-grained characterization of the
program's temporal behavior.  This may be useful for purposes other
than asymptotic complexity.  However, it means that the counting is
exquisitely sensitive to how the program is written.  For example, our
Python implementation of binary search contains the following if
statement.
\begin{lstlisting}
        if key == lst[mid]:
            return mid
        elif key < lst[mid]:
            high = mid - 1
        else:
            low = mid + 1
\end{lstlisting}
In this version, assuming that {\tt key} is not found initially, the
switch to upper half or lower half take exactly the same number of
steps. Assume instead that this were coded (as it was initially) as
follows:
\begin{lstlisting}
        if key < lst[mid]:
            high = mid - 1
        elif key == lst[mid]:
            return mid
        else:
            low = mid + 1
\end{lstlisting}
In this case, selecting the upper half of the list takes more steps
than selecting the lower half, since there's an additional test to
evaluate.  This means that the overall step count is dependent on the
{\it sequence} of choices made rather than the {\it number} of choices
made, as in the prior version.  This adds complexity to the overall
proof, even though the programs are functionally identical.  This
isn't too surprising since the exact timing of any program depends on
how it's written.  But in this case and many others, a difference
orthogonal to the big-O behavior substantially complicates the proofs.

A related issue is whether or not the code is optimized.  Consider the
following Python code:
\begin{lstlisting}
       while test:
          x = 0
          ...
\end{lstlisting}
Any sensible compiler would move the assignment to {\tt x} out of the
loop.  The step counts could vary significantly depending on whether
or not an optimization is performed, and could alter the big-O
behavior of a program.  We reasoned about the program as written
without regard to optimization.  It will be unsurprising that
properties of programs---both functional and
complexity-related---depend on software tools such as
compilers.\cite{BevierHuntMooreYoung89} However, these sorts of
subtleties make our proofs more fragile and significantly more labor
intensive than we would like.



\section{Conclusions and Future Work}
\label{Conclusions}

We have developed techniques for proving big-O properties of
imperative programs using ACL2.  This involves embedding a simple
imperative language in ACL2 with semantics defined with a standard
clocked interpreter.  We use the interpreter semantics to prove the
functional correctness of our code using standard techniques.  Our
interpreter also returns a count of the number of steps in the
execution of the program.  We use this count to prove that the number
of steps is bounded by a certain function in the big-O sense.  We
illustrated this with a proof that a certain program implements binary
search on a sorted list.  We show that the program is $O(\log_2(n))$,
where $n$ is the length of the list.  Because we analyze both the
functional behavior of a program and its step count concurrently, we
derive strong assurance that the program is both correct and is a
member of the expected big-O class.

However, reasoning about both functional and complexity results
imposes a pretty significant overhead if the primary goal is a big-O
result.  It might be possible to bound the program behavior without
carefully counting each step.  For binary search, for example, we
might prove that the recursive algorithm can't make more than
$\log_2(n)$ recursive calls on a list of length $n$.  ACL2 should be
an excellent tool for carrying out such a proof.  However, if the goal
is to prove complexity with respect to iterative implementions, we
would still like to verify that the recursive version computes the
same results as our iterative version.

Also, our current approach is highly sensitive to the way in which the
program is written and to what optimizations have been performed.
Overall, we have found our approach to proving big-O properties of
programs to be rather fragile and proof intensive.  We hope to find
approaches which require less effort and alleviate some of these
issues.

As noted by one of the anonymous reviewers, {\it clock functions} are
often used in ACL2 total correctness proofs relating to what Ray and
Moore\cite{RayMoore} call ``operationally modeled programs.''  Such
clock functions specify the number of steps required to reach a
halting state in programs for ``specialized architectures and machine
codes.''\cite{RayMoore} The reviewer suggested taking the ``clock
function used for functional correctness and then prov[ing] properties
of that clock function.''  We take this suggestion to mean that,
rather than developing a new imperative language and proving
properties about programs as we have done, to reason in terms of the
clock function for some of the several architectures or languages
already defined in ACL2.  There's no obvious reason to believe that
wouldn't work.  However, we're not convinced, as the reviewer suggests
that ``the method would have been applicable for any operational
semantics defined in ACL2 and [wouldn't] require tweaking and writing
a different evaluator.''  Unless the method were defined in
generality, it would still be specific to some particular architecture
or language, such as Moore's Java model\cite{Moore99}.  It might be
possible to define a {\it generic} language interpreter and prove
complexity properties for programs in that language, but that seems to
us more of the same.


We consider the research described in this paper to be preliminary.
In particular, it would be great to find more robust and automated
techniques for finding values for the existentially quantified
variables $c$ and $n_0$ from the Big-O definitions.  Another
potentially productive step would be constructing tools to aid in the
process of upper-bounding the number of steps of a computation.  The
approach described in this paper is labor intensive.  We regard this
work as a proof of principle that ACL2 can be used in developing
proofs of computational complexity.  But there's likely an easier way.


\nocite{*}
\bibliographystyle{eptcs}
\bibliography{biblio}
\end{document}